\documentclass{ws-mpla}
      \newcommand\arctanh{ {\, \rm arctanh \, } }
\begin{document}

\markboth{Turimov $\&$ Ahmedov} {{Electromagnetic Fields of
Rotating Magnetized Gravastars}}

\catchline{}{}{}{}{}

\title{Electromagnetic Fields of Slowly Rotating Magnetized Gravastars}

\author{\footnotesize B.~V.~Turimov}

\address{Ulugh Beg Astronomical Institute, Astronomicheskaya 33,
    Tashkent 100052, Uzbekistan \\
    bobur@astrin.uzsci.net}

\author{B.~J.~Ahmedov}

\address{Institute of Nuclear Physics, Ulughbek, Tashkent 100214, Uzbekistan\\
    Ulugh Beg Astronomical Institute, Astronomicheskaya 33,
    Tashkent 100052, Uzbekistan \\
    The Abdus Salam International Centre
for Theoretical Physics, 34014 Trieste, Italy\\
ahmedov@astrin.uzsci.net}

\author{A.~A.~Abdujabbarov}

\address{Institute of Nuclear Physics, Ulughbek, Tashkent 100214, Uzbekistan\\
    Ulugh Beg Astronomical Institute, Astronomicheskaya 33,
    Tashkent 100052, Uzbekistan \\
abahmadjon@yahoo.com}

\maketitle

\pub{Received (Day Month Year)}{Revised (Day Month Year)}

\begin{abstract}
We study the dipolar magnetic field configuration and present
solutions of Maxwell equations in the internal background
spacetime of a a slowly rotating gravastar. The shell of gravastar
where magnetic field penetrated is modeled as sphere consisting of
perfect highly magnetized fluid with infinite conductivity.
Dipolar magnetic field of the gravastar is produced by a circular
current loop symmetrically placed at radius $a$ at the equatorial
plane.

\end{abstract}

\ccode{PACS Nos.: 04.40.-b, 04.40.Dg, 04.40.Nr, 95.36.+x,
97.10.-q}

The existence of strong electromagnetic fields is one of the most
important features of rotating compact stars observed as pulsars
and magnetars with surface magnetic field as high as
$10^{14}G$~\cite{duncan}. On other hand the notion of a compact
astrophysical object has been one of the central issues in
contemplating general relativity as the theory of astrophysical
processes and structures. The search for solutions of Einstein
equations with different inputs for the physical sources of
gravitational field  has been one of the stepping stones on the
way to achieving comprehensible picture of the universe. Among
solutions found so far one eminent position certainly belongs to
black hole solutions with intriguing properties and
characteristics.

 The recent discovery of accelerating universe had inspired
 discussion of the existence of dark energy (see, for example, \cite{de,de1,de2,de3})
 and in turn
research for alternative configurations which led to a solution
dubbed gravastar, the gravitational vacuum star of dark energy, by
Mazur and Mottola~\cite{Mazur,Mazur2}. These spherically symmetric
static global solutions to the Einstein equations — candidates for
highly compact astrophysical objects and in this sense
alternatives to black holes — evolve from the segment of the de
Sitter geometry in the center with the equation of state of dark
energy, proceed through a thin vacuum phase transition layer,
avoid formation of the event horizon, and swiftly match the
exterior Schwarzschild spacetime. The common feature of gravastar
realization is the anisotropy of pressure in the outer shell of
the object.

Several astrophysically relevant aspects of gravastar solutions
such as thermodynamic properties, modes of quasi-normal
oscillations, and ergoregion instability were recently discussed
in the literature. De Benedictis et al~\cite{benedictis} and
Chirenti and Rezzolla~\cite{cecilia,cecilia2} investigated the
stability of the original model of Mazur and Mottola against axial
perturbations, and found that gravastars are stable to these
perturbations. Chirenti and Rezzolla also showed that their
quasi-normal modes differ from those of a black hole of the same
mass, and thus can be used to distinguish a gravastar from a black
hole. The aim of this short note is to find interior
electromagnetic fields in the shell of gravastar consisting of
perfect highly magnetized fluid with infinite conductivity. It is
assumed that dipolar magnetic field of the gravastar is produced
by a circular current loop symmetrically placed at radius $a$ at
the equatorial plane.

Throughout, we use a space-like signature $(-,+,+,+)$ and a system
of units in which $G = 1 = c$ (However, for those expressions with
an astrophysical application we have written the speed of light
explicitly.). Greek indices are taken to run from 0 to 3 and Latin
indices from 1 to 3; covariant derivatives are denoted with a
semi-colon and partial derivatives with a comma.

Metric, which is describing spacetime the spherical symmetric
slowly rotating gravastar, can be written in the following form
(see, for example,~\cite{Mazur,Mazur2,Rocha}):
\begin{equation}
\label{gvs}
ds^{2}=-A^{2}(r)dt^{2}+A^{-2}(r)dr^{2}+r^{2}d\theta^{2}+r^{2}\sin^{2}\theta
d\varphi^{2}-2\omega(r) r^{2}\sin^{2}\theta d\varphi dt
\end{equation}
where $\emph{r}$  is the radial coordinate, $\omega(r)$ is the
angular velocity of dragging of inertial frames around slowly
rotating gravastar, and here
$$
{A^{2}(r)}=\left\{
  \begin{array}{lcr}
   1-\frac{2M}{r},\qquad r>a(\tau),\\
   \\
   1-\frac{r^{2}}{R^{2}},\qquad r<a(\tau).
  \end{array}
  \right.
$$
where $r = a(\tau)$ is a timelike hypersurface, at which the
infinitely thin shell is located, and $\tau$ denotes the proper
time of the thin shell and the constant $R=\sqrt{a^3/2M}$.

A circular current loop with current $I$ and net charge $q$,
symmetrically placed at radius $a$ in the equatorial plane of a
slowly rotating gravastar, has electric current density components
as in~\cite{pet74}:
\begin{eqnarray}
\label{cura} && J^{\hat{t}}=\frac{qA}{2\pi
a^2}\delta(r-a)\delta(\cos\theta)  \ ,
\\
\label{curb} && J^{\hat{r}}=J^{\hat{\theta}}=0 \ ,
\\
\label{curd} && J^{\hat{\varphi}}=\frac{Ir\sin\theta}{2\pi
a^2}\delta(r-a)\delta(\cos\theta)\ ,
\end{eqnarray}
where 'hatted' quantities are the orthonormal components measured
by zero angular momentum observers (ZAMO) with four-velocity
\begin{equation}
(u^{\alpha})_{\rm ZAMO}
\equiv\frac{1}{A}\left(1,0,0,\omega\right),\hskip 2.0cm
(u_{\alpha})_{\rm ZAMO}\equiv A\left(-1,0,0,0\right)\ .
\end{equation}

Gravastar is slowly rotating with 4-velocity:
\begin{equation}
w^{\alpha}\equiv \frac{1}{A}\left(1,0,0,\Omega\right),\hskip 2.0cm
w_{\alpha}\equiv A\left(-1,0,0,\frac{\bar{\omega}
r^{2}sin^{2}\theta}{A^{2}}\right)\ ,
\end{equation}
where $\bar{\omega}=\Omega-\omega$, $\Omega$ is angular velocity
of rotation of gravastar.

 We now
look for an interior solution of Maxwell equations in background
spacetime given by (\ref{gvs}) assuming that magnetic field of the
star is dipolar. To simply the search for a solution we look for
separable solutions of Maxwell equations in the form
\begin{eqnarray}
\label{ansatz_1} && B^{\hat r}(r,\theta) = F(r)\cos\theta\ ,
\\\nonumber\\
\label{ansatz_2} && B^{\hat \theta}(r,\theta) = G(r)\sin\theta\ ,
\\\nonumber\\
\label{ansatz_3} && B^{\hat \phi}(r,\theta) = 0\ ,
\end{eqnarray}
\noindent where functions $F(r)$ and $G(r)$ will account for the
relativistic corrections due to a curved background spacetime
~\cite{ram01}.

    Maxwell equations  with the ansatz
(\ref{ansatz_1})--(\ref{ansatz_3}), yield the following set of
equations
\begin{eqnarray}
\label{ir_1} &&\left(r^2 F\right)_{, r} + 2 r G/A = 0 \ ,
\\\nonumber\\
\label{ir_2} &&\left(r A G\right)_{, r}    +  F = 0\ .
\end{eqnarray}

\noindent Note a first important result in the system of equations
(\ref{ir_1})--(\ref{ir_2}). In the case of stationary
electromagnetic fields, the general relativistic frame dragging
effect and gravitomagnetic charge do not introduce a correction to
the radial eigenfunctions of the magnetic fields. In other words,
in the case of infinite conductivity and as far as the magnetic
field is concerned, the study of Maxwell equations in a slow
rotation metric provides no additional information with respect to
a non-rotating metric. The dependence from the frame dragging
effects is therefore expected to appear at ${\mathcal
O}(\omega^2)$.

The stationary vacuum magnetic field external to an aligned
magnetized relativistic star is well known and given
by~\cite{go64}
\begin{eqnarray}\label{hamjac}
 && B_{\rm ex}^{\hat r}(r,\theta) =-\frac{3\mu}{4M^3}
    \left[\ln N^2 + \frac{2M}{r}\left(1 +  \frac{M}{r}
    \right)\right]\cos\theta
    \ ,
\\\nonumber\\
\label{sol_mfe_2} && B_{\rm ex}^{\hat \theta}(r,\theta) =
\frac{3\mu N}{4 M^2 r}
    \left[\frac{r}{M}\ln N^2 +\frac{1}{N^2}+ 1
    \right] \sin\theta\ ,
\end{eqnarray}
where $\mu=\pi a^2\left(1-2M/a\right)^{1/2}I$ and
$N=\left(1-2M/r\right)^{1/2}$ is lapse function.

\begin{figure*}
\includegraphics[width=0.5\textwidth]{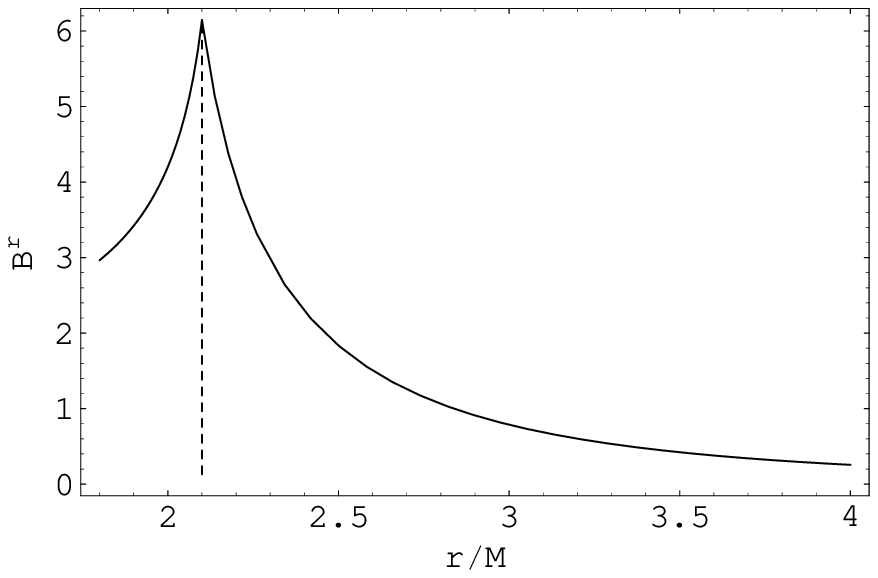}%
\includegraphics[width=0.5\textwidth]{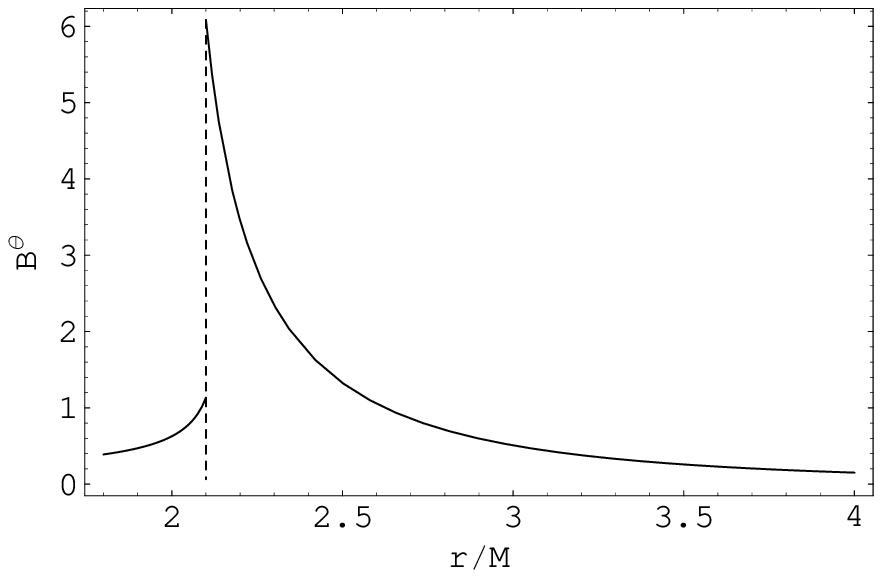}
\caption{\label{fig:1} Dependence of the radial (left) and
azimuthal (right) components of the magnetic field of the
gravastar from the radius. Interior magnetic field inside of the
shell increases up to the border at $a$ (here $a/M=2.1$) which is
indicated with the vertical dash line. Azimuthal component of
magnetic field subjects to the discontinuity across the boundary
of gravastar. Exterior magnetic field of gravastar is decaying as
$1/r^3$.}
\end{figure*}

Interior solution for magnetic field is
\begin{eqnarray}\label{hamjac1}
\label{sol_mfe_1}  B^{\hat r}(r,\theta) &=& -\frac{3\mu a^{3}}
{4r^3M^3} \left[\ln
N_{a}^{2}+\frac{2M}{a}\left(1+\frac{M}{a}\right)\right]
\frac{\frac{r}{R}-\arctanh \frac{r}{R}}{\frac{a}{R}-\arctanh
\frac{a}{R}}\cos\theta
    \ ,
\\\nonumber\\
\label{sol_mfe_2_new} B^{\hat \theta}(r,\theta) &=&\frac{3\mu a^2L
N_{a}} {4r^3L_{a}M^2}\left[\frac{a}{M}\ln
N^2_{a}+\frac{1}{3N_a^2}-\frac{2N_a^2}{3}+\frac{7}{3}\right]
\frac{\frac{r}{RL^2}- \arctanh\frac{r}{R}}
{\frac{a}{RL_{a}^2}-\arctanh\frac{a}{R}} \sin\theta,\qquad
\end{eqnarray}
where $L=\left(1-r^{2}/R^{2}\right)^{1/2}$, subscript $a$ denotes
quantities measured at $r=a$. The values of the integration
constant are defined from the continuity of the radial magnetic
field across the gravastar surface, i.e. that
$[B^{\hat{r}}_a]_{in}=[B^{\hat{r}}_a]_{ext}$, boundary condition
for the tangential magnetic field
$[B^{\hat{\theta}}_a]_{ext}-[B^{\hat{\theta}}_a]_{in}=4\pi
i^{\hat{\varphi}}_a$, where $\mathbf{i}$ is density of the surface
current at $r=a$. The radial dependence of magnetic field of
gravastar (\ref{hamjac})--(\ref{sol_mfe_2_new}) is plotted in
Fig.~\ref{fig:1}.

Interior electric fields
\begin{eqnarray}
\label{sol_ef_1} && E^{\hat
r}(r,\theta)=\frac{\bar{\omega}r\sin\theta}{cL}B^{\hat
\theta}(r,\theta)
    \ ,
\\\nonumber\\
\label{sol_ef_2} && E^{\hat \theta}(r,\theta)
=-\frac{\bar{\omega}r\sin\theta}{cL}B^{\hat r}(r,\theta) .
\end{eqnarray}
can be found from the condition of the infinite conductivity
$\sigma$ in Ohm law with the explicit components of the conduction
current $j^{\hat \alpha}$ as
\begin{eqnarray}
\label{current2} &&j^{\hat r} =  \sigma
        \left[ E^{\hat r} +A^{-1}\left( v^{\hat\theta} B^{\hat\varphi}
        - v^{\hat\varphi}B^{\hat \theta}\right)\right] \ ,
\\\nonumber\\
\label{current3} &&j^{\hat\theta} = \sigma\left[E^{\hat \theta}+
        A^{-1}\left( v^{\hat\varphi} B^{\hat r}
        - v^{\hat r}B^{\hat \varphi}\right)\right] \ ,
\\\nonumber\\
\label{current4} &&j^{\hat\varphi} = \sigma \left[E^{\hat \varphi}
+
        A^{-1}\left( v^{\hat r} B^{\hat\theta}
        - v^{\hat \theta}B^{\hat r}\right)\right] \ ,
\end{eqnarray}
and using expressions for velocity of rotation $v^{\hat\varphi}$
and magnetic field (\ref{sol_mfe_1}), (\ref{sol_mfe_2_new}).

Exterior electric field of slowly rotating gravastar coincides
with the electric field of rotating neutron star and given by
equations (124)--(126) in the paper~\cite{ram01}.

In this paper we have derived components of the dipolar magnetic
field of the gravastar which is produced by a circular current
loop symmetrically placed at radius $a$ at the equatorial plane.
The knowledge of gravastar's magnetic field can be useful for
description of different physical processes in the gravastar.

\section*{Acknowledgments}

Authors thank the IUCAA for warm hospitality during their stay in
Pune and AS-ICTP for the travel support through BIPTUN (NET-53)
program. This research is supported in part by the UzFFR (projects
5-08 and 29-08) and projects FA-F2-F079 and FA-F2-F061 of the UzAS
and by the ICTP through the OEA-PRJ-29 project and the Regular
Associateship grant.

\end{document}